\magnification=\magstep1 
\baselineskip=24pt 
\input amssym.def 
\input amssym.tex 
\overfullrule0pt 
\centerline{\bf HYDRODYNAMICAL EQUATION FOR ELECTRON SWARMS} 
\centerline {Joel L. Lebowitz} 
\centerline{Department of Mathematics and Physics}
\centerline{Rutgers University} 
\centerline{Piscataway, New Jersey 08854-8019} 
\centerline{e-mail: lebowitz@math.rutgers.edu}
\centerline{and} 
\centerline{A. Rokhlenko} 
\centerline{Department of Mathematics} 
\centerline{Rutgers University} 
\centerline{Piscataway, New Jersey 08854-8019} 
\centerline{e-mail: rokhlenk@math.rutgers.edu}

\bigskip
\noindent
{\bf Abstract}

We study the long time behavior of light particles, e.g.\ an electron
swarm in which Coulomb interactions are unimportant, subjected to an
external field and elastic collisions with an inert neutral gas.  The
time evolution of the velocity and position distribution function is
described by a linear Boltzmann equation (LBE).  The small ratio of
electron to neutral masses, $\epsilon$, makes the energy transfer
between them very inefficient.  We show that under suitable scalings
the LBE reduces, in the limit $\epsilon \to 0$, to a formally exact
equation for the speed (energy) and position distribution of the
electrons which contains mixed spatial and speed derivatives. When the
system is spatially homogeneous this equation reduces to and thus
justifies, for $\epsilon$ small enough, the commonly used ``two-term''
approximation.
\bigskip
\noindent
{\bf Introduction}

The motion of an electron under the influence of an external electric
field and elastic ``collisions'' with the ``background'' is a problem
of both historic and contemporary interest. It arises in gas
discharges in the laboratory and in the atmosphere, when the degree of
ionization is so low that electron-electron and electron-ion
interactions are negligible, the so called swarm approximation [Carron
1992, Allis 1956, Franklin 1976, Margenau and Hartman 1948].  Similar
situations arise in the study of electron transport in solids in a
semi-classical regime [Lorentz 1916, Sommerfeld 1956, Ben Abdolllah
and Degon 1996, Ben Abdolllah {\it et al} 1996, Golse and Poupaud
1992].  In such situations it is often reasonable to treat the
scatterers as if they were in thermal equilibrium.  {}For electrons
moving in a (almost) neutral gas with constant density $\rho$ this
corresponds to the atoms having a Maxwellian velocity distribution
$F({\bf V}; M)$ at a specified temperature $T,$
$$
F({\bf V};M) = \left({2\pi kT\over M}\right )^{-d/2}
\exp\left (-{M V^2\over 2kT}\right ), \eqno(1)
$$
where $M$ is the mass of the scatterers, e.g.\ the neutral atoms in a
noble gas, $k$ is the Boltzmann constant, and $d$ is the space dimension.

The time evolution of the electron distribution function is then given
by a linear Boltzmann equation (LBE) of the form [Franklin 1976,
Landau and Lifshits 1993],
$$
{\partial f({\bf r},{\bf v},t) \over \partial t} + {\bf v} \cdot {\partial
\over \partial {\bf r}} f - {e{\bf E} \over m} \cdot {\partial \over 
\partial {\bf
v}} f = \int[K({\bf v}, {\bf v}^\prime) f({\bf r}, {\bf v}^\prime, t) -
K({\bf v}^\prime, {\bf v}) f({\bf r}, {\bf v}, t)]d{\bf v}^\prime =
{\cal K}f, \eqno(2)
$$
where $e$, $m$ are the charge and mass of an electron, {\bf E} is the
external field and $K$ gives the rate for scattering.  The transition
rate $K$ will depend on the nature of the scatterers but, for the case
of elastic collisions, to which we restrict ourselves here, $K$ will
satisfy detailed balance with respect to a Maxwellian velocity
distribution for the electrons $F({\bf v};m)$, at the same temperature
as that of the neutrals, so that $ K({\bf v}, {\bf v}^\prime) F({\bf
v}^\prime; m) = K({\bf v}^\prime, {\bf v}) F({\bf v}; m)$.  This
ensures that for $E=0$, $F({\bf v};m)$ is a stationary velocity
distribution which will be approached by $f({\bf v},t)$ as $t \to
\infty$; $f({\bf v},t)$ is obtained from $f({\bf r}, {\bf v},t)$ by
integrating over the spatial coordinate $\bf r$.  We assume for
simplicity that the electrons are confined to a large periodic box:
$f$ is then a probability density with $\int \int f({\bf r}, {\bf
v};t) d{\bf r} d{\bf v} = 1$.  It is also possible to consider other
boundary conditions as well as electric and magnetic fields which are
space and time dependent, but we shall not do so here. We shall also
not consider here explicitly electrons in solids for which the
scatterers are various kind of excitations [Ben Abdolllah and Degon
1996, Ben Abdolllah {\it et al} 1996, Golse and Poupaud 1992].

The central physical fact about the electron-neutral system is the
great disparity in their masses: $ {m\over M}=\epsilon <
10^{-4}$. Consequently the change in the speed of an electron in a
typical collision, which is of order $\sqrt{\epsilon}$, will be very
small, while the change in the velocity direction will be large,
becoming independent of $\epsilon$ when $\epsilon
\to 0$ [Franklin 1976, Landau and Lifshits 1993].  This suggests
approximating the integral operator in (2) by expanding the right side
of (2) in powers of $\epsilon$ and dropping ``small terms'' in
$\epsilon$.  There is a certain amount of ambiguity in carrying out
such an expansion, arising from the uncertainty of how to treat the
dependence of the unknown $f$ on $\epsilon$.  A reasonable choice
gives
$$
{\partial f \over \partial t} + {\bf v} \cdot {\partial f \over \partial
{\bf r}} - {e{\bf E}\over m} \cdot {\partial f \over \partial {\bf v}} = 
{\epsilon \rho\over v^2}{\partial \over \partial v}
\left [{v^4\sigma (v)}\left (f_0+{kT\over mv}{\partial f_0
\over \partial v}\right )\right ]+{v\rho\sigma (v)}(f_0-f).\eqno(3)
$$
In (3) $v = |{\bf v}|$, $f_0({\bf r},v,t)$ is the average of $f({\bf
r},{\bf v},t)$ over angles and $\sigma(v)$ is the collision cross
section, see [Margenau and Hartman 1948, Ginzburg and Gurevich 1960].
The dependence of $f$ on $\epsilon$ is then to be determined from the
solution of (3).

There is, however, very little control over this expansion.  In fact
eq.(3) is not necessarily positivity preserving.  Nevertheless eq.(3)
yields reasonable answers for the stationary velocity distribution of
the electrons, ${\bar f}({\bf v})$.  The latter can be obtained
explicitly in the so-called ``two-term'' approximation [Lorentz 1916,
Margenau and Hartman 1948, Allis 1956, Carron 1991], in which one
keeps only the first two terms in a spherical harmonic expansion of
$\bar f({\bf v})$.  This distribution was first found by Druyvesteyn
[Druyvesteyn 1930, Druyvesteyn and Penning 1940], who considered the
case $T=0,\ \sigma =const$, and was later generalized by Davydov
[Davydov 1935] for all $T$ and $\sigma$. {}For an analysis of the error
made by the two term approximation to (3) see [Rokhlenko and Lebowitz
1997, Rokhlenko - submitted].

A somewhat different approach to this problem was taken by Koura
[Koura 1987]. Starting with a two term $\epsilon$-expanded kinetic
equation (slightly different than what is obtained from (3)) he
observed, that after a scaling of space, time and electric field and
neglecting the time variation of the first harmonic, one is led to an
equation in which $\epsilon$ does not appear at all.  Koura then
argued that the actual physical quantities of interest in an electron
swarm should have a similar simple scaling dependence on $\epsilon$,
when $\epsilon$ is small.  This has the advantage of permitting more
efficient computer simulations at larger than realistic value of
$\epsilon$. Doing simulations for several values of $\epsilon$ Koura
found good agreement with results from the scaled two term
approximation for $\epsilon ^{<}\hskip-1.5ex_{\sim}10^{-2}$.

In this note we shall use the same scaling as Koura but apply it
directly to (2) without any other approximation. This is in the same
spirit as the scalings used for electrons in solids [Ben Abdolllah and
Degon 1996, Ben Abdolllah {\it et al} 1996, Golse and Poupaud 1992].
It is based on a formulation of space-time scalings now commonly used
to obtain a ``reduced hydrodynamic description'' from a more detailed
microscopic one, see [Spohn 1995].

\bigskip\noindent
{\bf One-dimensional problem}

To simplify matters we shall first consider the one dimensional
version of (2), corresponds to hard collisions of point particles with
masses $m$ and $M=m/\epsilon$,
$$
{\partial f(x,{\bf v},t)\over \partial t}+{\bf v}{\partial f(x,{\bf v},t)
\over \partial
x}-{eE\over m}{\partial f(x,{\bf v},t)\over \partial {\bf v}}=$$
\vskip-.6truein
$$
~~~~\eqno(4)
$$
\vskip-.6truein
$$\rho \int_{-\infty}^{\infty}|{\bf v}-V|\left \{f\left
[x, {2V-(1-\epsilon){\bf v}\over 1+\epsilon},t\right ]F\left
[{(1-\epsilon)V + 2\epsilon {\bf v} \over 1 + \epsilon};{m\over \epsilon}
\right ] - f(x,{\bf v},t)F(V;m/\epsilon ) \right \}dV,
$$
where ${\bf v}\in {\Bbb R}$. Eq. (5) is to be 
solved subject to some initial condition $f(x,{\bf v},0)$.  

To obtain the behavior of $f$ for long times we rescale our variables
by setting $y = \sqrt \epsilon x$, $\tau = \epsilon t$ (diffusive
scaling [Spohn 1995]) and $E = \sqrt \epsilon E^*$ (the field has to
be small on this scale for the energy to remain bounded when the time
is of order $\epsilon^{-1}$). We now define the even and odd parts of
the velocity distribution function
$$
\phi_\epsilon(y,v,\tau)= {1 \over 2 \sqrt \epsilon}[f(y/\sqrt
\epsilon,v,\tau/\epsilon) + f(y/\sqrt \epsilon,-v, \tau/\epsilon)],
$$
\vskip-.6truein
$$
~~~~\eqno(5)
$$
\vskip-.6truein
$$
\sqrt \epsilon \psi_\epsilon(y,v,\tau) = {1 \over 2\sqrt \epsilon}
[f(y/\sqrt \epsilon,v,t/\epsilon) -
f(y/\sqrt \epsilon,-v,t/\epsilon)],
$$
where $v = |{\bf v}|$ and we have put $f \to {1 \over \sqrt \epsilon}
f$ to preserve the normalization: in the scaled variable $y$ the
system is in a periodic box of size $L$ independent of $\epsilon$.  By
changing the integration variable in (3) we then obtain two coupled
equations for $\phi_\epsilon$ and $\psi_\epsilon$, $$ {\partial
\phi_\epsilon (y,v,\tau) \over \partial \tau} + v{\partial
\psi_\epsilon\over\partial y} - {eE^*\over m} {\partial
\psi_\epsilon \over \partial v} = \eqno(6)
$$
$$
\epsilon^{-1} \rho\sqrt{m\over 2\pi kT}\int_{-\infty}^{\infty} 
|v-V|\left \{\left ({1+\epsilon\over 1-\epsilon }\right )^2 
\phi_\epsilon \left [y, {(1+\epsilon)v -
2V \over 1-\epsilon},\tau\right ] - \phi_\epsilon(y,v,\tau )\right \}
e^{-{m V^2\over 2\epsilon kT}}dV,
$$
\vskip-.6truein
$$
~~~~
$$
\vskip-.6truein
$$
\epsilon {\partial \psi_\epsilon (y,v,\tau )\over \partial \tau} + 
v{\partial \psi_\epsilon\over \partial y} - {eE^*\over m}
{\partial \phi_\epsilon \over \partial v} =
$$
$$ 
-\rho\sqrt{m\over 2\pi kT}\int_{-\infty}^{\infty} 
|v-V|\left \{\left ({1+\epsilon\over 1-\epsilon }\right )^2 
\psi_\epsilon \left [y, {(1+\epsilon)v -
2V \over 1-\epsilon},\tau\right ] + \psi_\epsilon(y,v,\tau )\right \}
e^{-{m V^2\over 2\epsilon kT}}dV.$$

We assume now that the initial distribution is such that
$\phi_\epsilon$ and $\psi_\epsilon$ have well defined limits
$\phi(y,v,0)$ and $\psi(y,v,0)$ as $\epsilon \to 0$.  Taking now
formally the limit $\epsilon \to 0$ on both sides of (7), keeping
$\tau ,y$ and $E^*$ fixed we get the limiting equations,
$$
{\partial \phi(y,v,\tau) \over \partial \tau} + v{\partial \psi \over
\partial y} - {eE^* \over m}{\partial \psi
\over \partial v} = 2\rho {\partial \over \partial v}\left [v^2\left
(\phi + {kT \over
m}{\partial \phi \over \partial v}\right )\right ],\eqno(8a)
$$
$$
v {\partial \phi \over \partial y} - {eE^* \over m}{\partial \phi
\over \partial v} = -2\rho v\psi(y,v,\tau ),\eqno(8b)$$
valid for $v\geq 0.$ Solving (8b) for $\psi$ and substituting into
(8a) we get the reduced equation for the even part of the distribution
$\phi(y,v,\tau)$, in terms of the scaled space, time and electric
field
$$
{\partial \phi \over \partial \tau} + {1 \over \rho} {\partial
\over \partial y}\left ({eE^* \over m} {\partial \phi \over \partial v} -
{v\over 2}{\partial \phi \over \partial y}\right ) = {\partial \over 
\partial v}
\left\{\left [\left ({eE^* \over m}\right )^2 {1 \over 2\rho v} + 
{kT \over m}2\rho v\right ]{\partial
\phi \over \partial v} + 2\rho v^2 \phi\right \},\eqno(9)
$$
with the boundary condition at $v=0$, $$
{\partial \phi (y,0,\tau )\over \partial y}={eE^*\over m}\lim_{v\to +0}
{1\over v}{\partial \phi (y,v,\tau )\over \partial v}\eqno(10)$$
and the normalization 
$$\int_0^L dy \int_0^{\infty} \phi(y,v,\tau)dv = 1/2.$$ 
The condition (10) at $v=0$ follows from the (assumed) continuity of
the distribution $f$ at ${\bf v}=0$.

Let us introduce the rescaled electron density $n(y,\tau)$, drift
$u(y,\tau)$, and mean speed $w(y,\tau)$:
$$
n(y,\tau )=2 \int_0^\infty \phi(y,v,\tau)dv,$$
\vskip-.7truein
$$\eqno(11)$$
\vskip-.7truein
$$
n(y,\tau) u(y,\tau) = 2\int_0^\infty v \psi(y,v,\tau )dv,
\ \ n(y,\tau ) w(y,\tau ) = 2\int_0^{\infty} v \phi(y,v,
\tau )dv.$$
Integrating Eq.\ (9) over velocities, we obtain the continuity
equations  
$$
{\partial n(y,\tau ) \over \partial \tau } + {\partial (nu)\over 
\partial y} =0.\eqno(12)
$$

To obtain $u$ we substitute (8b) into (11) and integrate over $v$, 
$$
u =-{e\phi (y,0,\tau )\over m\rho n(y,\tau )} E^*- {w(y,\tau ) \over 
2\rho n(y,\tau )}{\partial n \over \partial y}- {1\over 2\rho
}{\partial w(y,\tau )\over 
\partial y}. \eqno(13)
$$
One can identify in (13) the factors in front of $E^* ,\ {\partial n
\over\partial y},\ {\partial w \over\partial y}$ as respectively the
mobility, diffusion coefficient, and a parameter related to
thermodiffusivity [Golant {\it et al.} 1980].

The right side of eq.(9) coresponds to a drift-diffusion in ``speed
space'' with a diffusion coefficient given by the term in the square
brackets and with a drift $-2 \rho v^2$.  The spatially hompgeneous
stationary solution of (9) has the form 
$$
\bar \phi (v)=C \exp\left [-\int_0^v {s^3ds\over {kT\over m}s^2+\left ({eE^*
\over 2m\rho}\right )^2}\right ],\eqno(14)
$$
similar to the Davydov distribution for $d=3$, c.f.\ eq.\ (21).  Note that
for $E^*=0$, i.e. no external field, $\bar \phi (v)=F(v;m)$ as it should.

\bigskip\noindent
\noindent {\bf Three dimensions}

In $d=3$ the right side of the LBE takes the form 
$$
{\cal K}f = \rho \int  \int |{\bf v} - {\bf V}|\sigma({\bf v} -
{\bf V})\left[ f({\bf v}^\prime,t) F({\bf V}^\prime;M) - f({\bf v},t)
F({\bf V};M)\right] d{\bf V} d\hat \omega
$$
The angular integration is over the scattering solid angle $\hat
\omega$ of the electron in the center of mass coordinate system,
$\sigma ({\bf v})$ is the electron-neutral differential collision
cross section, and we have
$$
{\bf v}^\prime = {\bf v} + {2 \over 1+\epsilon} \hat \omega[\hat \omega
\cdot({\bf V}-{\bf v})],
$$
$$
{\bf V}^\prime = {\bf V} - {2\epsilon \over 1+\epsilon} \hat \omega[\hat
\omega \cdot ({\bf V} - {\bf v})].
$$

Carrying out a similar analysis as for $d = 1$ we separate $f$ into a
spherically symmetric part $\phi_{\epsilon} ({\bf y},v,\tau )$ and a
remainder $\psi_{\epsilon}({\bf y},{\bf v},\tau )$, i.e.\ we set
$$f({\bf y}/\sqrt{\epsilon},{\bf v}, t/\epsilon )=
\epsilon^{3/2}[\phi_{\epsilon} ({\bf y},v,\tau )+\sqrt{\epsilon}
\psi_{\epsilon} ({\bf y,v},\tau )],\eqno(15)
$$
$$
\phi_{\epsilon}({\bf y},v,\tau )={\epsilon^{-3/2}\over 4\pi}\int 
f({\bf y}/\sqrt{\epsilon},{\bf v}, t/\epsilon )d\Omega.\eqno(16)
$$
In (16) the integration is over the unit sphere specifying the
orientation of the vector ${{\bf v} \over v}$.  Taking now formally
the limit $\epsilon\to 0$ we obtain, in terms of the rescaled
variables, a
set of equations for $\phi$ and $\psi$ entirely analogous to (8),(9),
$$
{\partial\phi\over \partial\tau}-{v\over 3\rho\sigma}{\bf \nabla}^2_y\phi+
{e\over 3m\rho}{\bf E^*}\cdot\nabla_y\left [{1\over \sigma}
{\partial\phi\over \partial v}+{1\over v^2}{\partial\over \partial v}
\left ({v^2\phi\over\sigma }\right )\right ]=$$
\vskip-0.6truein
$$\eqno(17)$$
\vskip-0.6truein
$${1\over v^2}{\partial \over 
\partial v}\left [\left ({eE^*\over m}\right )^2{v\over 3\sigma \rho} 
{\partial \phi\over \partial v}+\rho v^4\sigma (v)\left (\phi +{kT
\over mv}{\partial\phi\over \partial v}\right )\right ],$$
$$\psi = {1\over \rho \sigma (v)v}\left ({e\over m}\bf E^*\cdot
\nabla_v\phi-{\bf v}\cdot \nabla_y\phi\right ).\eqno(18)$$
We assume the collisions to be spherically symmetric, so $\sigma (v)=
4\pi \sigma ({\bf v})$ is the total cross section. 

The spatial density and current are given by$$
n({\bf y},\tau )=4\pi\int_0^\infty v^2\phi({\bf y},v,\tau )dv,
\ \ n({\bf y},\tau) {\bf u}({\bf y},\tau).\eqno(19)$$
They satisfy equation of continuity$$
{\partial n({\bf y},\tau )\over \partial\tau}+\nabla_y[n{\bf u}] =
0.\eqno(20)$$
The electron drift ${\bf u}$ has the form
$$
{\bf u} = \mu {\bf E^*} - D{1 \over n} {\bf \nabla}_y n - {1 \over 3\rho}
{\bf \nabla}_y \left\langle {v \over \sigma(v)} \right\rangle,
$$
$$
\mu = {e \over 3\rho m} \left\langle {1 \over v^2}{d \over dv}\left [{v^2 
\over\sigma(v)}\right ]\right\rangle, \quad \quad 
D = {1 \over 3\rho} \left\langle {v \over \sigma(v)} \right\rangle,
$$
where $\langle g(v) \rangle = {4\pi \over n} \int_0^\infty g v^2
\phi({\bf y},v,\tau)dv$ is the average over the velocity distribution.  
In the spatially homogeneous case eq. (17) coincides with the usual
equation obtained in the two term approximation, whose stationary
solution is
$$
\phi (v)=C\exp{\left [-\int_0^v {s^3ds\over {kT\over m}s^2+{1\over 3}
\left ({eE^*\over m\rho \sigma}\right )^2}\right ]},\eqno(21)$$
which was given by Davydov in [Davydov 1935].  {}For $E^* = 0$ $\phi(v)$
in (21) is just $F({\bf v};m)$ while for $T=0,\ \sigma = const$ it
coincides with the Druyvesteyn distribution.
\bigskip\noindent 
{\bf Inequalities for moments at arbitrary $\epsilon$ and T=0}

In previous works [Rokhlenko and Lebowitz 1997, Rokhlenko - submitted]
we established bounds on the first few moments of the stationary
distribution function using the approximate equation (3) with a
velocity independent total cross section $\sigma_0$.  Here we shall do
the same for the LBE (2) when $T=0.$ Introducing the dimensionless
variables,$$ x={{\bf v\cdot E}\over vE},\ {\bf
s=v}\epsilon^{1/4}\sqrt{m\rho\sigma_0
\over eE},$$
and noting that for $T=0$, $F({\bf V};M) = \delta({\bf V})$, the
stationary, spatially homogeneous, $f$ satisfies, in $d=3$, the
equation,
$$
-\sqrt{\epsilon}\left (x{\partial
f\over \partial s}+{1-x^2\over s}{\partial f\over \partial x}\right )=$$
\vskip-0.7truein
$$\eqno(22)$$
\vskip-0.7truein
$${(1+\epsilon )^2 \over 2\pi}\int f({\bf s'}) \delta
[(1-\epsilon )(s')^2+2\epsilon ({\bf s\cdot s'})-(1+\epsilon )s^2]d{\bf s'}
-s f({\bf s}).$$

{}Following [Lorentz 1916, Margenau and Hartman 1948, Rokhlenko 1991] we
expand the distribution function in Legendre series$$
f(s,x)=\sum_{l=0}^{\infty}f_l(s)P_l(x),$$ substitute it into (22) and
obtain a set of coupled equations for the $f_l$. Let us introduce the
moments
$$
M_l(k)=\int_0^{\infty}s^kf_l(s)ds.\eqno(23)$$ We can now use the same
technique as in [Rokhlenko and Lebowitz 1997] to obtain bounds on the
mean energy $W(\epsilon )$ and drift $u(\epsilon )$ which are defined
in terms of $M_0(k).$ These inequalities are based on the fact that
$\log{M_0(k)}$ is a convex function of $k.$ This yield in the present
case
$$
{\sqrt{b}\over a}\leq{2\rho \sigma\sqrt{\epsilon}\over eE}W(\epsilon )
\leq{1\over \sqrt{a}},
\ \ \ {b^{1/4}\over a}\leq {2\epsilon^{3/4}\over |p_0(3,\epsilon )|}
\sqrt{m\rho\sigma_0\over eE}u(\epsilon ) \leq {1\over
a^{3/4}},\eqno(24)
$$
where
$$ a={p_0(4,\epsilon)p_1(2,\epsilon)
\over 3\epsilon },\ \ b={p_0(7,\epsilon)p_1(5,\epsilon)\over 12
\epsilon [1+p_0(3,\epsilon)/p_2(3,\epsilon)]}\eqno(25)$$
and $$
p_l(k,\epsilon )={(1+\epsilon )^2\over 2\epsilon }\int^1_{|1-\epsilon/
1+\epsilon |}
x^kP_l\left [{(1+\epsilon)x^2+\epsilon -1\over 2\epsilon x}\right ]dx
-1.\eqno(26)$$

When $\epsilon \to \infty$ the lower and upper bounds for $u$ and $W$
merge giving in the limit$$ u=\sqrt{eE\over M\rho\sigma_0},\ \
W={mu^2\over 2}.\eqno(27)$$ When $\epsilon \to 0$, $a\to 1,\ \ b\to
1/2$ and the inequalities (24) are then satisfied by the Druyvesteyn
distribution. We believe that with greater effort it should be
possible to obtain upper and lower bounds which when $\epsilon \to 0$
both converge to the values obtained from the Davydov distribution.

\noindent {\bf Acknowledgments}

The research was supported by NSF Grant No. 95-23266 and AFOSR Grant
No. 95-0159.  
\vfill\eject
\centerline {\bf REFERENCES}
\bigskip\noindent
W.P.Allis, {\it Handb.Phys.} {\bf 21}, 383 (1956)
\medskip\noindent
N.Ben Abdollah and P.Degond, {\it J.Math.Phys.} {\bf 37}, 3306 (1996)
\medskip\noindent
N.Ben Abdollah, P.Degond, and S. Genies, {\it J.Stat.Phys.} {\bf 84},
205 (1996)
\medskip\noindent
N.J.Carron, {\it Phys.Rev.A} {\bf 45}, 2499-2511 (1991).
\medskip\noindent
P.Davydov, {\it Phys. Z. Sowjetunion} {\bf 8}, 59-70 (1935).
\medskip\noindent
M.J.Druyvesteyn, {\it Physica} {\bf 10}, 61 (1930)
\medskip\noindent
M.J.Druyvesteyn and E.M.Penning, {\it Rev.Mod.Phys.} {\bf 12}, 87 (1940)
\medskip\noindent
R.Esposito, J.L.Lebowitz, and R.Marra, {\it J.Stat.Phys.} {\bf 78},
389 (1995)
\medskip\noindent
R.N.Franklin, {\it Plasma Phenomena in Gas Discharges} (Clarendon, Oxford, 
1976)
\medskip\noindent
V.L.Ginzburg and A.V.Gurevich, {\it Sov.Phys.-Usp.} {\bf 3}, 115 (1960)
\medskip\noindent
V.E.Golant, A.P.Zhylinsky, and I.E.Sakharov, {\it 
Fundamentals of Plasma Physics} (Wiley, New York, 1980)
\medskip\noindent
F.Golse and F.Poupaud, {\it Asympt.Anal.} {\bf 6}, 135 (1992)
\medskip\noindent
K.Koura, {\it J.Phys.Soc.Japan} {\bf 56}, 429-432 (1987)
\medskip\noindent
L.D.Landau and E.M.Lifshitz, {\it Physical Kinetics} 
(Pergamon Press, New York, 1993)
\medskip\noindent
H.A.Lorentz, {\it The Theory of Electrons} (B.G.Taubner
Leipzig, 1916)
\medskip\noindent
H.Margenau and L.M.Hartman, {\it Phys.Rev.} {\bf 73}, 309-315 (1948)
\medskip\noindent
A.Rokhlenko, {\it Phys.Rev.A} {\bf 43}, 4438-4451 (1991)
\medskip\noindent
A.Rokhlenko and J.L.Lebowitz, {\it Phys.Rev.E}, {\bf 56}, 1012-
1018 (1997)
\medskip\noindent
A.Rokhlenko, {\it Phys.Rev.E} (submitted and processed)
\medskip\noindent
I.P.Shkarofsky, T.W.Johnston, and M.P.Bachynski, {\it The Particle
Kinetics of Plasma} (Addison-Wesley, Reading MA, 1966) 
\medskip\noindent
A.Sommerfeld, {\it Thermodynamics and Statistical
Mechanics} (Acad. Press, New York, 1956)
\medskip\noindent
H.Spohn, {\it Large Scale Dynamics of Interacting
Particles} (Springen, 1995)

\end